\documentclass[american,aip,rsi,reprint]{revtex4-1}
\usepackage{babel}
\usepackage{graphicx}
\usepackage{dcolumn}
\usepackage{bm}
\usepackage{subcaption}
\usepackage{natbib}

\begin{document}
\title{\large Design and development of a low temperature, inductance based high frequency ac susceptometer}
\author{E. Riordan}
\affiliation{Department of Physics \& Astronomy, Cardiff University, Cardiff, CF24 3AA, United Kingdom}
\author{J. Blomgren}
\author{C. Jonasson}
\author{F. Ahrentorp}
\author{C. Johansson}
\affiliation{RISE, Arvid Hedvalls Backe 4, Box 53071, SE-400, G\"{o}teborg, Sweden}
\author{D. Margineda}
\affiliation{Department of Physics \& Astronomy, Cardiff University, Cardiff, CF24 3AA, United Kingdom}
\author{A. Elfassi}
\author{S. Michel}
\author{F. Dell'ova}
\affiliation{INSA, Institut National des Sciences Appliquées, TOULOUSE, 135 Avenue de Rangueil, 31077 Toulouse Cedex 4, France}
\author{G. M. Klemencic}
\author{S. R. Giblin}
\affiliation{Department of Physics \& Astronomy, Cardiff University, Cardiff, CF24 3AA, United Kingdom}

\begin{abstract}
	We report on the development of an induction based low temperature high frequency ac susceptometer capable of measuring  at frequencies up to 3.5 MHz and at temperatures between 2 K and 300 K. Careful balancing of the detection coils and calibration have allowed a sample magnetic moment resolution of $5\times10^{-10} Am^2$ at 1 MHz. We will discuss the design and characterization of the susceptometer, and explain the calibration process. We also include some example measurements on the spin ice material CdEr$_2$S$_4$ and iron oxide based nanoparticles to illustrate functionality.
\end{abstract}
\maketitle

\section{Introduction}
Ac susceptibility is a common technique for the characterization of magnetic materials, in which a small oscillating magnetic field, \(H\), is applied to the sample and the differential change in the magnetization, \(M\), of the sample allows the susceptibility, \(dM/dH\), to be measured. The ac susceptibility can be represented by the complex susceptibility: $\chi = \chi' + i\chi''$. The in-phase component $\chi'$ represents the slope of the $M(H)$ curve at low frequency and is in-phase with the excitation ac-field. The out-of-phase component $\chi''$ (with phase at 90 degrees with respect to the excitation field) represents energy losses from the excitation field to the sample~\cite{Martien1994}. One such process is magnetic relaxation effects, where the out-of-phase component of the susceptibility exhibits a peak at a frequency corresponding to a characteristic relaxation time. Magnetic relaxation dependent processes include the intrinsic N\'{e}el relaxation and Brownian relaxation which can demonstrate extrinsic properties~\cite{Bogren2015}.

Induction based ac susceptometers capable of measuring at low temperatures are generally limited to the kHz range; those that do exist and allow measurement into the MHz range are currently limited to room temperature~\cite{Ahrentorp2010}. 
Other techniques have been used to measure ac susceptibility at low temperatures and high frequencies, including the use of resonant circuits~\cite{Dahlberg1979} and the tunnel diode oscillator technique~\cite{clover1970} which require long measurement times for multiple frequencies. Toroidal based systems are capable of measuring into the GHz range but are limited by low sensitivity and difficulties in mounting of the sample~\cite{Grambow1971,Fannin1986,Hanson1991}. Micro-SQUID systems, suitable for microscopic samples have also worked at high frequencies~\cite{Boyd2009}. The system discussed in this paper is a versatile induction based system allowing both the in-phase and out-of-phase components of the susceptibility to be easily separated up to the MHz regime for bulk samples; moreover the system works in commercial cryostats allowing temperatures from 2-300 K and dc magnetic fields to be swept from 0-9 T. This allows access to relaxation effects in samples across a varied range of phase space previously only accessible by central facility techniques such as muon spin rotation which require a particle accelerator. This system will have useful applications in many areas of physics including frustrated magnetism, superconductivity and magnetic nanoparticles which can be used for biomedical applications. 

An induction based ac susceptometer consists of two circuits, an excitation circuit and detection circuit. In a first order gradiometer (such as our susceptometer) the detection circuit consists of two in-series counter wound coils connected to a lock-in amplifier. The detection coils are counter wound to ensure a minimal response to the application of any external field. More complex second order coils can be designed however for systems at high frequency the inductive and capacitive response is of more concern.
During an ac susceptibility measurement a sinusoidally oscillating magnetic field is applied to the sample by the excitation coil. The sample is initially placed in the center of one of the detection coils, changing the magnetic flux and therefore changing the voltage detected by the lock-in amplifier. The sample is then moved to the center of the second detection coil causing a voltage change of the opposite sign. The difference between the measured voltage in the upper and lower coils can then be related to the susceptibility by
\begin{equation}
\Delta V = V_U - V_L = \mu_0 \omega \alpha N A  H_0  (\chi''+i\chi') 
\end{equation}

\noindent where, $V_U$ and $V_L$ are measured voltages when the sample is in the upper and lower detection coil respectively, $N$ is the number of turns in the detection coils, $A$ is the cross sectional area of the coils, $\omega$ is the angular frequency being measured ($\omega=2\pi f$ where $f$ is the excitation frequency), $H_0$ is the amplitude of the excitation field, $\alpha$ is a coupling factor that is dependent on the coil and sample geometries, $\chi'$ and $\chi''$ are the in-phase and out-of-phase components of the susceptibility respectively, and $\mu_0$ is the permeability of free space. To find $\alpha$ is essential and is usually experimentally verified by calibrating the system with a sample that has a known in- and out-of-phase component of its frequency dependent ac susceptibility; historically Dy$_2$O$_3$ meets the requirement for the frequency range of interest~\cite{Chen2011}.


A major difficulty in achieving high frequencies in induction based ac susceptometers is to reduce the parasitic capacitance and inductance due to the windings of the excitation and detection coils. This is necessary for two reasons. Firstly, the parasitic capacitance and inductance give a resonance at a specific frequency that must be well above the measured frequency range. Secondly, the field produced by the excitation coil is directly proportional to the current through it, so if the impedance (due to coil reactance) increases then the field amplitude decreases, a smaller field means a smaller detectable moment for a given sample susceptibility. 

\section{Coil set \& interface}
Our susceptometer is similar to the Dynomag HF system~\cite{Ahrentorp2010}, which allows measurements up to 10 MHz at room temperature. The ac susceptometer presented in this paper consists of a NbTi excitation coil with $N$=24 turns, wound in a double layered helix so as to minimize parasitic capacitance. Two $N$=2 turn shim coils are used at both ends to achieve a field that is uniform along the axis of the excitation coil. NbTi was chosen to reduce parasitic heating below the superconducting transition ($T_{c}=9.8 K$). The detection coils consist of two in-series counter-wound copper coils of N=15 turns each.

Measurement of the excitation and detection coil resistances, inductances and LC-resonances, together with LTSpice analysis, have shown a good frequency stability up to 5~MHz. A high Q-value resonance in the detection coil is prone to induce oscillations in the detection circuit. To reduce the Q-value it is common practice to put damping resistances in parallel with each counter-wound detection coil in the detection circuit. In the current setup the detection coil is connected directly to a lock-in amplifier and hence the damping can alternatively be achieved by using a 50~$\Omega$ input impedance on the lock-in amplifier which eliminates the need for damping resistors. In practice our system is capable of measuring up to 3.5~MHz, however the minimum detectable ac susceptibility increases above 1.5~MHz due to the increasing coil impedance reducing the excitation current as the excitation circuit begins to approach resonance.

\begin{figure}[!ht]
\centering
\includegraphics[width=\linewidth,keepaspectratio]{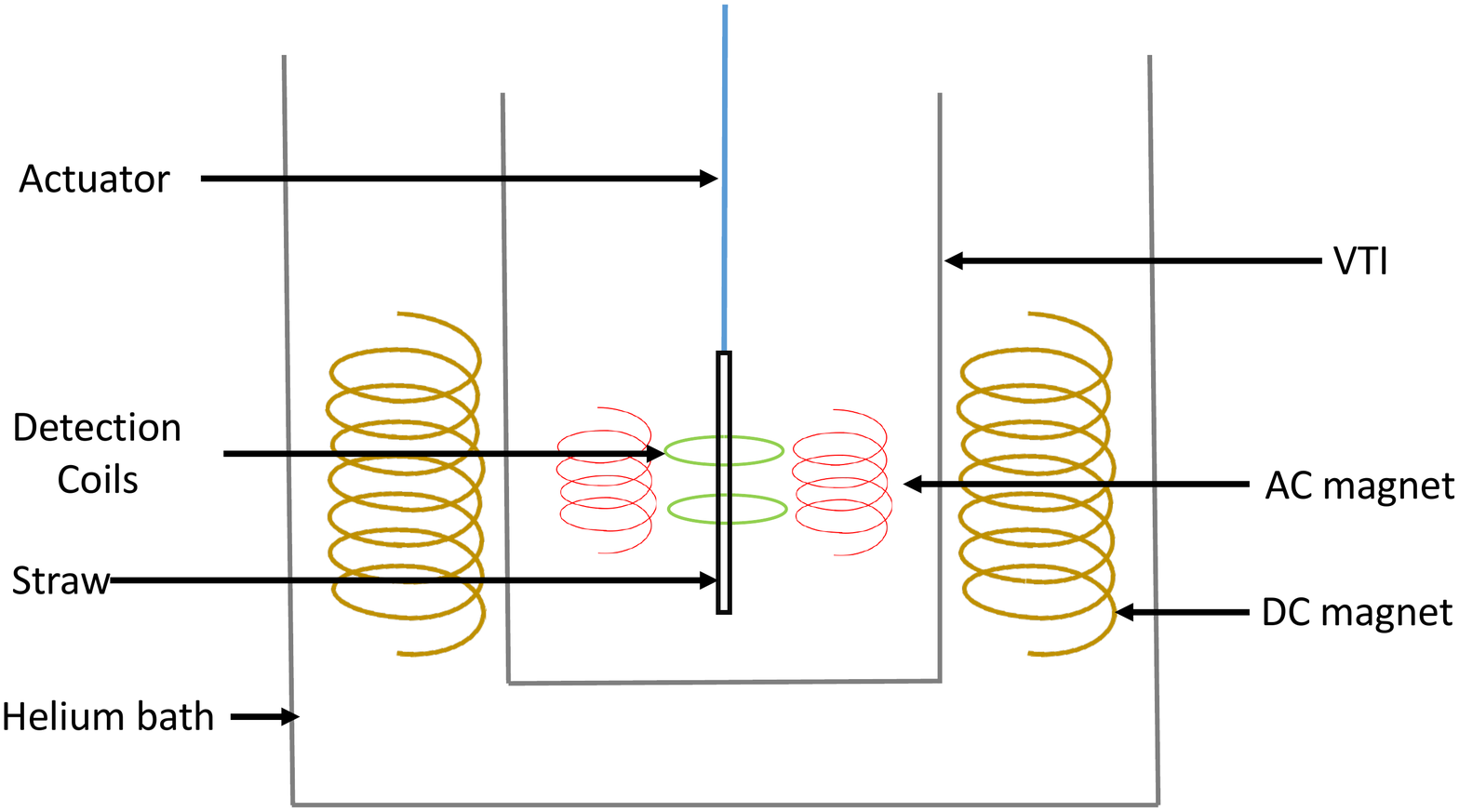}
\caption{\label{diagram} \textit{Sketch of the experimental system including the Helium bath cryostat and the variable temperature insert (VTI). The DC magnet is in the Helium bath with the AC magnet and detection coils in the VTI.}}
\end{figure}

Our ac susceptometer is designed for use in a Quantum Design physical property measurement system (QD-PPMS). Integration with a QD-PPMS via a Labview interface enables measurement between 2~K and 300~K and in magnetic fields up to 9~T. Our system has been extensively tested in externally applied static fields up to 3~T at temperatures down to 2 K. Although the temperature stability of the system relies upon gas exchange with the sample, careful calibration is required to ensure no parasitic heating; this can be measured by knowing the temperature dependent response of known samples and careful monitoring and calibration of the system with thermometers in the heat exchanger and the neck of the variable temperature insert. The neck thermometer in the wall of the sample chamber also allows the ability to monitor effective eddy current heating in the walls from the excitation field. A Stanford Research Systems DS345 signal generator is used to drive the excitation coil and Stanford Research Systems SR844 or SR830 lock-in amplifiers (the SR830 with a 50 $\Omega$ terminator) are used for the phase sensitive detection. The SR844 is capable of measuring well beyond the upper frequency limit of our system, but limits the lowest accessible frequency to 25~kHz. The SR830 can measure well below the lower limit of our system, but limits the maximum frequency to 102~kHz. The sample is moved between the detection coils using an Accu-glass 2'' linear actuator mounted at the top of the sample space. Samples are mounted in plastic drinking straws which fix to the end of a long carbon fiber rod that attaches to the actuator magnetically. Figure~\ref{diagram} shows an experimental sketch of the hardware.

\section{Excitation Coil Characterization}
\begin{figure}[!ht]

\centering
\includegraphics[width=\linewidth]{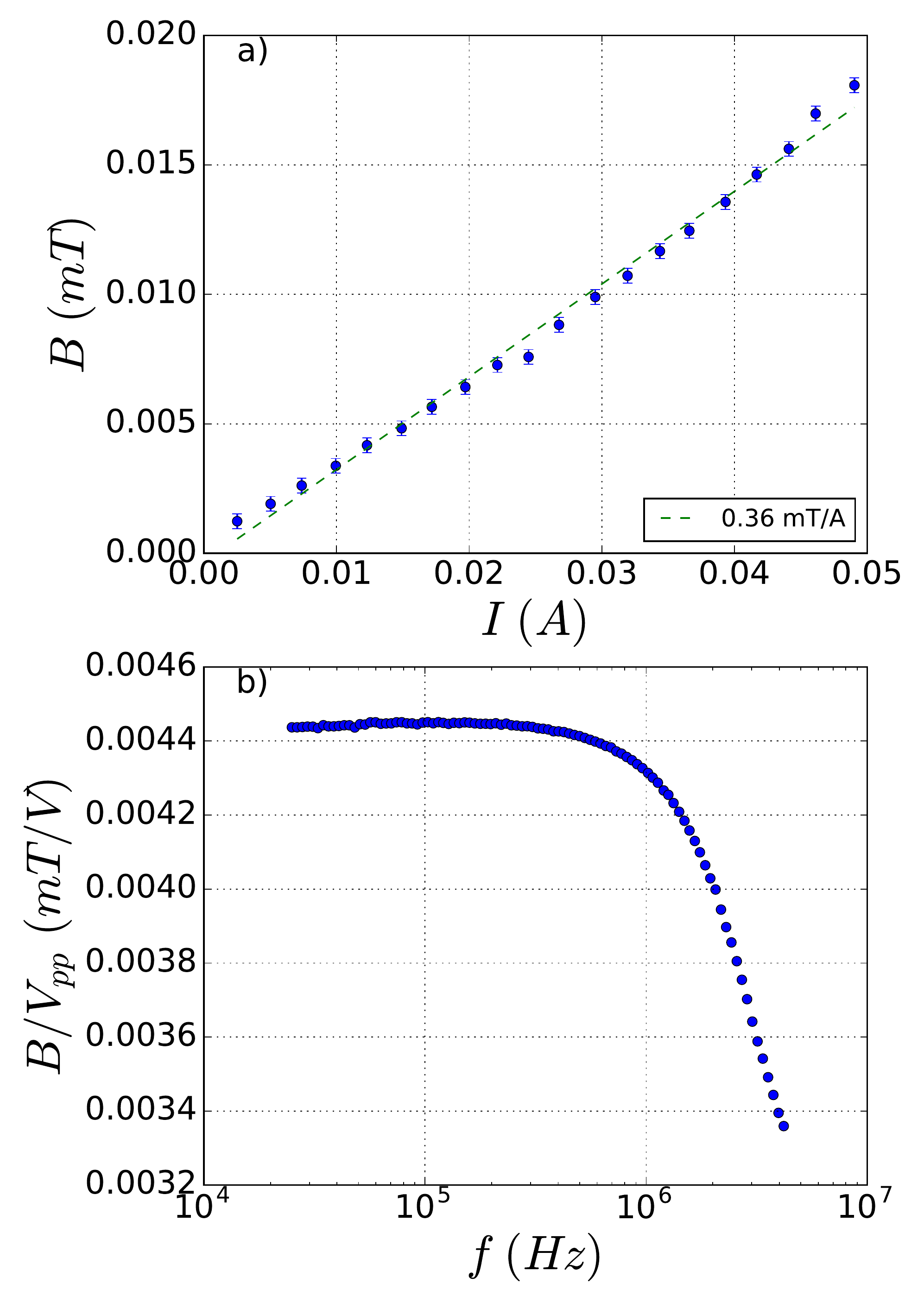}
\caption{\label{BVAFvsF}\textit{a) Magnetic flux in the center of the coil set as measured by a longitudinal hall probe against applied DC current, with the linear fit showing the calibration factor. b): Frequency dependence of the excitation field.}}
\end{figure}

The excitation field was calibrated using a longitudinal Hall probe with a DC current applied to the coil, from this the field produced per unit applied current (B/I) can be calculated. B/I  can then be used to measure the response of the coil as a function of frequency for an ac current. Figure~\ref{BVAFvsF}a displays the result of these measurements, resulting in a B/I value of 0.358$\pm$0.008 mT/A. The measured excitation field is in agreement with the fit within measurement error at large currents. The frequency characteristics of the current through the excitation coil were measured using a custom made analogue differential input amplifier (DIA) and an oscilloscope. It is then trivial to calculate how the field amplitude behaves with frequency, the results of which are shown in figure~\ref{BVAFvsF}b.  
A 90.9~$\Omega$ resistor is placed in series with the excitation coil when it is in the superconducting phase in order to achieve a known excitation field. It is also possible to remove this resistor to apply a much larger excitation field if required but that will be limited by the signal generator's 50~$\Omega$ series impedance. 


\section{Measurement Process \& Calibration}
\begin{figure}[!ht]
\centering
\includegraphics[width=.9\linewidth,keepaspectratio]{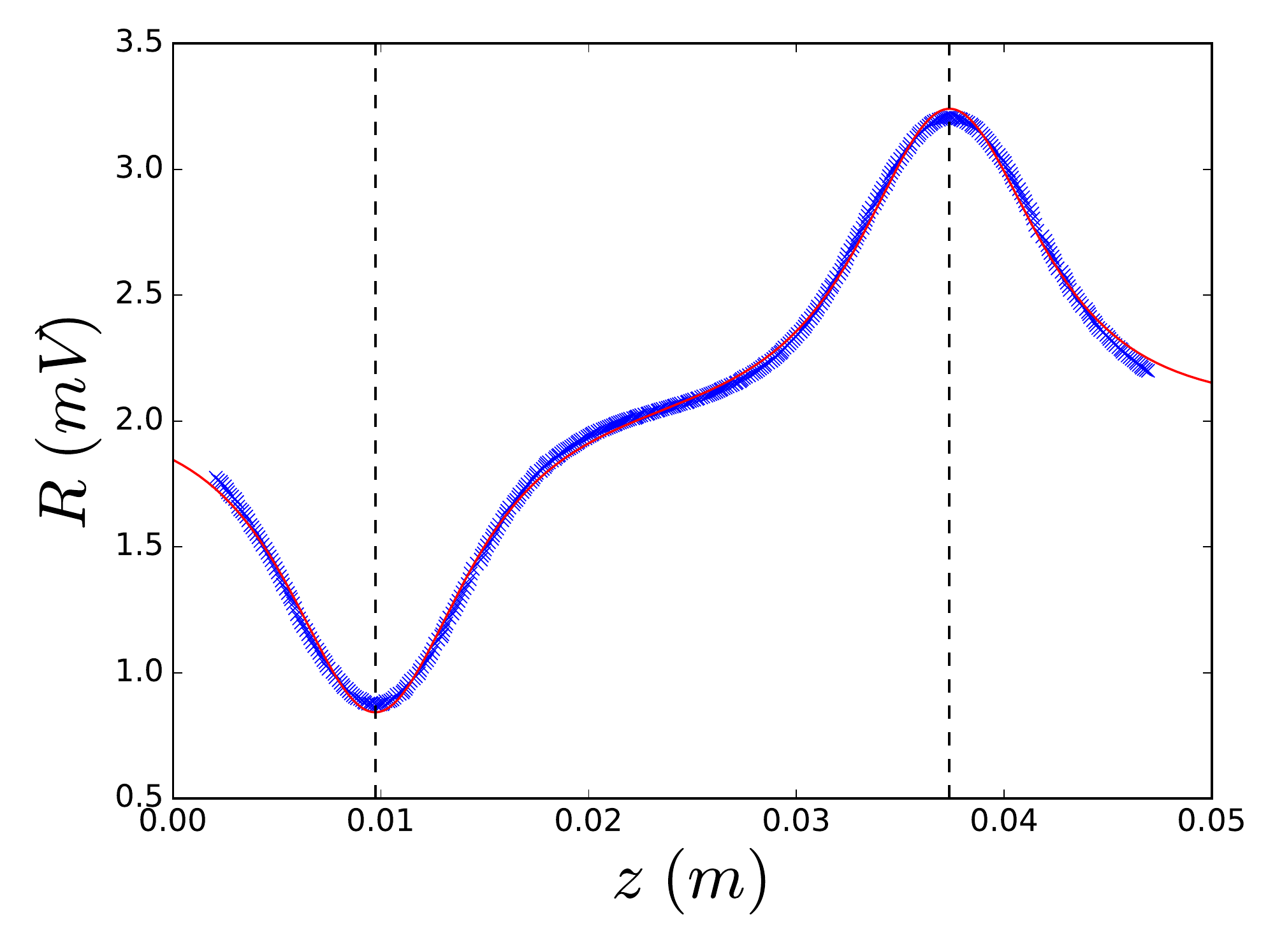}
\caption{\label{scancoil} \textit{Magnitude of the raw voltage recorded on the lock-in amplifier against the sample position in the coil set. Blue crosses show the raw data and the red line is a fit to equation 2. Vertical dashed lines indicate detection coil centers.}}
\end{figure}

The measurement of the susceptibility is performed by the Dynomag LTHF LabVIEW program \cite{RISE2017}, which is integrated with the QD-PPMS software. To perform any measurement with the susceptometer it is first necessary to locate the sample relative to the detection coils. To do this the sample is moved through the coils and the magnitude of the complex voltage, $R$, is recorded as a function of position. This trace is then fitted with an equation (for the case of a first-order gradiometer) of the form
\begin{equation}
R(z)=B\frac{{A_1}^2}{({A_1}^2+(z-z_1)^2)^{3/2}}-B\frac{{A_2}^2}{({A_2}^2+(z-z_2)^2)^{3/2}}+C
\end{equation}
\noindent where $A_1$, $A_2$, $B$, and $C$ are fitting constants. $z_1$ and $z_2$ are the positions of the upper and lower coils. An example trace and fit are shown in figure~\ref{scancoil} showing the coil locations at $z_1=0.0098$ m and $z_2=0.0374$ m. The position of the peak is an essential component in the calibration of the system as the absolute magnitude of the measured susceptibility depends upon correct placement of the measurement sample relative to the detection coils.

\begin{figure}[!ht]
\centering
\includegraphics[width=\linewidth,keepaspectratio]{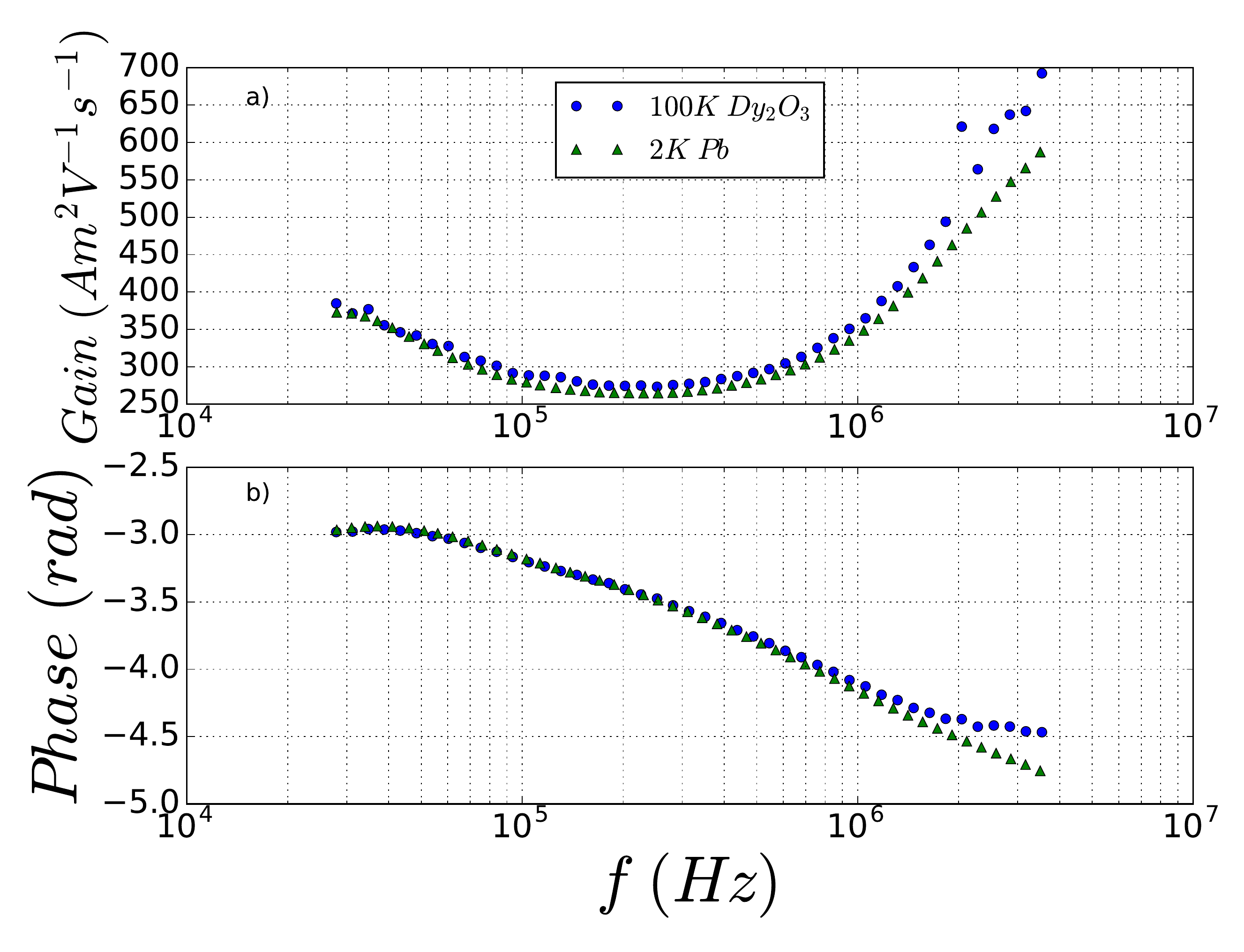}
\caption{\label{GPs}\textit{a) Example gain calibrations at different frequencies. b) Example of phase corrections at different frequencies. In both panels blue circles show a calibration taken at 100~K and the green triangles show calibrations at 2~K.}}
\end{figure}

Two further processes are required for the calibration. The first involves a simple background measurement, where an empty sample holder is moved between the detection coils and the voltages are measured in the upper and lower coils. The difference is subtracted from the voltages recorded during a measurement. The second, a so-called gain and phase calibration, is fundamental to the magnitude of the extracted value of the susceptibility. A gain and phase calibration is used to determine an appropriate scaling factor for the measured susceptibility and to correct for frequency dependent phase shifts introduced by the complex impedance of the equipment. Samples with purely real and well known susceptibilities are required and then the observed susceptibility is adjusted to match the known sensitivity. A key consideration of the system is the response over the entire temperature range. We have used two well understood samples to cover a large temperature range; Dy$_2$O$_3$, a paramagnetic material for the high temperature regime and Pb, a superconductor at low temperature. In both cases only a sample with a purely real and well known susceptibility is required. It is impractical to calibrate at every temperature and so calibrations are generally only performed at 2 K for the Pb and 100 K for Dy$_2$O$_3$ which are then used for measurements across a broad temperature range. Examples of gain and phase calibrations are shown in figure~\ref{GPs}a and b respectively across a range of temperature and frequency. Both calibration samples produce similar phase corrections (some difference is expected because the excitation coil is superconducting below 9.8 K) and near identical gain factors. This simple check of the system performance enables the temperature to be scanned with some confidence. 

\begin{figure}[!ht]
\centering
\includegraphics[width=.9\linewidth,keepaspectratio]{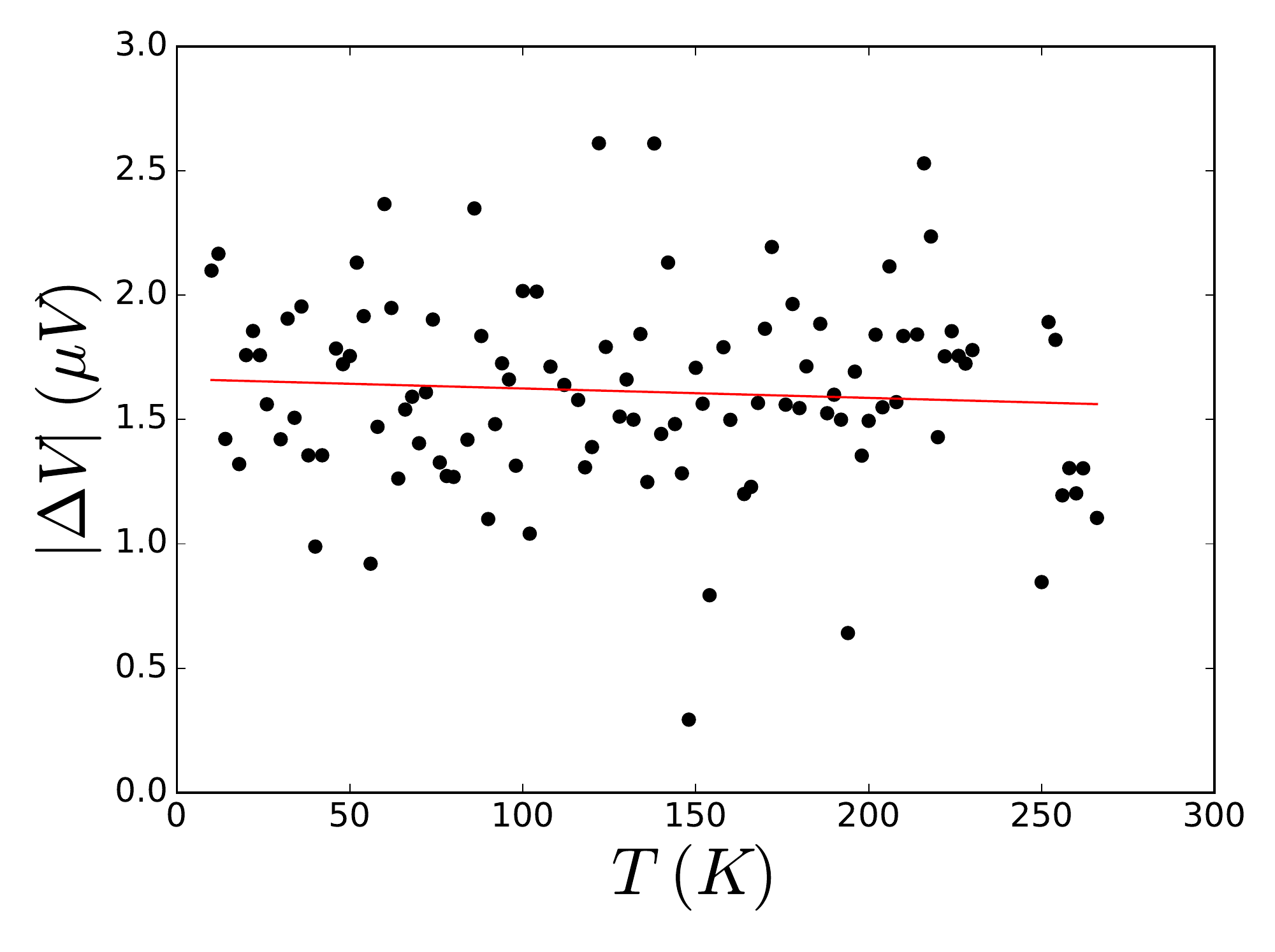}
\caption{\label{BLNK}\textit{Magnitude of the measured complex voltage against temperature, as detected by the lock-in for the case of an empty sample holder at 1~MHz. Red line is a linear fit to the data.}}
\end{figure}

An empty sample holder was measured across the full temperature range to check for an inherent temperature dependence of the system that would have to be accounted for in measurements. Calibration was performed as described above (at 100 K) and the background voltage amplitude was measured from 10K-266 K as shown in figure~\ref{BLNK}. The amplitude of the excitation field is known to vary according to the temperature dependence of the resistance of the NbTi wire and so this has been corrected for in figure~\ref{BLNK} by scaling the detected voltage appropriately. As shown in figure~\ref{BLNK}, a linear least squares fit has found a gradient of $(-3.79\pm 5.27)\times10^{-10}~V/K$ indicating a very weak to non existent relationship between temperature and detected voltage, as any change is within the background noise. The voltages shown in figure~\ref{BLNK} are two orders of magnitude smaller than typical measurement voltages. 



\section{Measurements}

\begin{figure}[!ht]
\centering
\includegraphics[width=\linewidth,keepaspectratio]{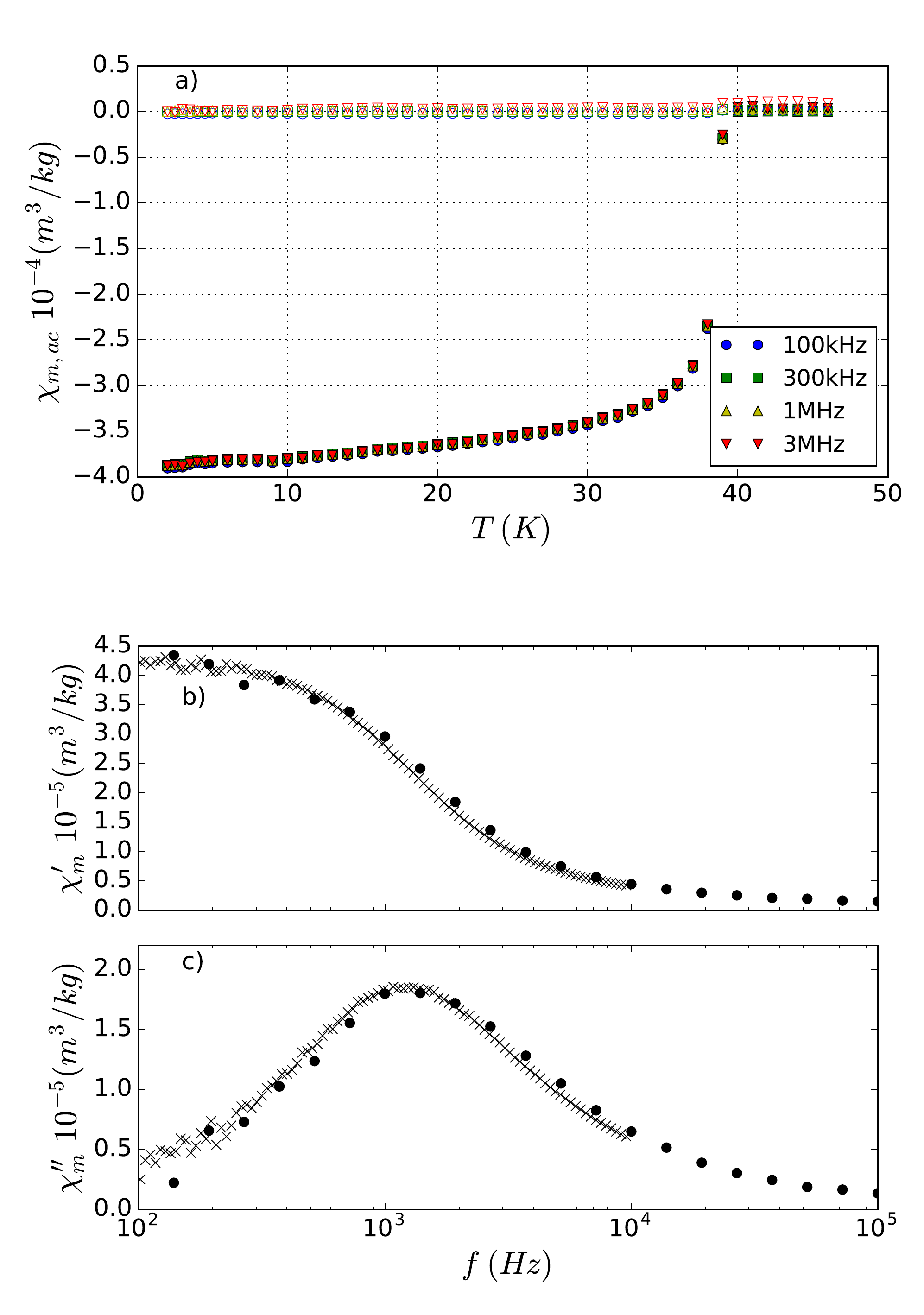}
\caption{\label{Pbmes} \textit{Data to illustrate temperature reliability. a) Mass ac-susceptibility vs. temperature on a powder MgB$_2$ sample showing the transition to the superconducting state at 39 K. Solid markers indicate the in-phase component and empty markers indicate the out-of-phase component. b) In-phase mass ac-susceptibility on a single crystal of Dy$_2$Ti$_2$O$_7$ at 15 K. Dots indicate data taken by our high frequency set up and crosses indicate data taken by the QD-ACMS susceptometer. c) Out of phase component part of b.}}
\end{figure}

To ensure system reliability it is essential to demonstrate there is no heating effect on the QD-PPMS by the coil. Therefore we have attempted to cover a broad temperature range at high frequency to ensure systematic behavior of the system by examining the behavior of known samples. We have measured the superconductor MgB$_2$, which has a superconducting critical temperature of 39~K, at a number of frequencies. Figure~\ref{Pbmes}a, shows the in-phase component of the diamagnetic susceptibility through the critical temperature, and the determined transition is in agreement with literature. No imaginary part is observed as at the high frequencies and low amplitude ($H_0$=0.045~mT) of our system the peak is expected to be too narrow for the temperature step used~\cite{Ciszek2010a} (1 K). A demagnetization correction has been applied such that $\chi_V '(T_{min})=-1$, where $\chi_V '(T)$ is the in-phase component of the samples volume ac-susceptibility at temperature $T$. Clearly the data are in agreement demonstrating the stability at high temperature with frequency. Furthermore, the frequency dependent properties are checked by considering a sample of the spin ice Dy$_2$Ti$_2$O$_7$; the in- and out-of-phase component shown in figure 6b and c respectively were measured at 15~K using the ac susceptibility option (QD-ACMS) of the QD-PPMS. We can check consistency of our system by overlaying the QD-ACMS data taken at similar frequencies to data taken using the coil set designed for this work and extend the data to higher frequency. The data show a clear overlap and demonstrate the accuracy of the equipment.


\begin{figure}[!ht]
\centering
\includegraphics[width=\linewidth,keepaspectratio]{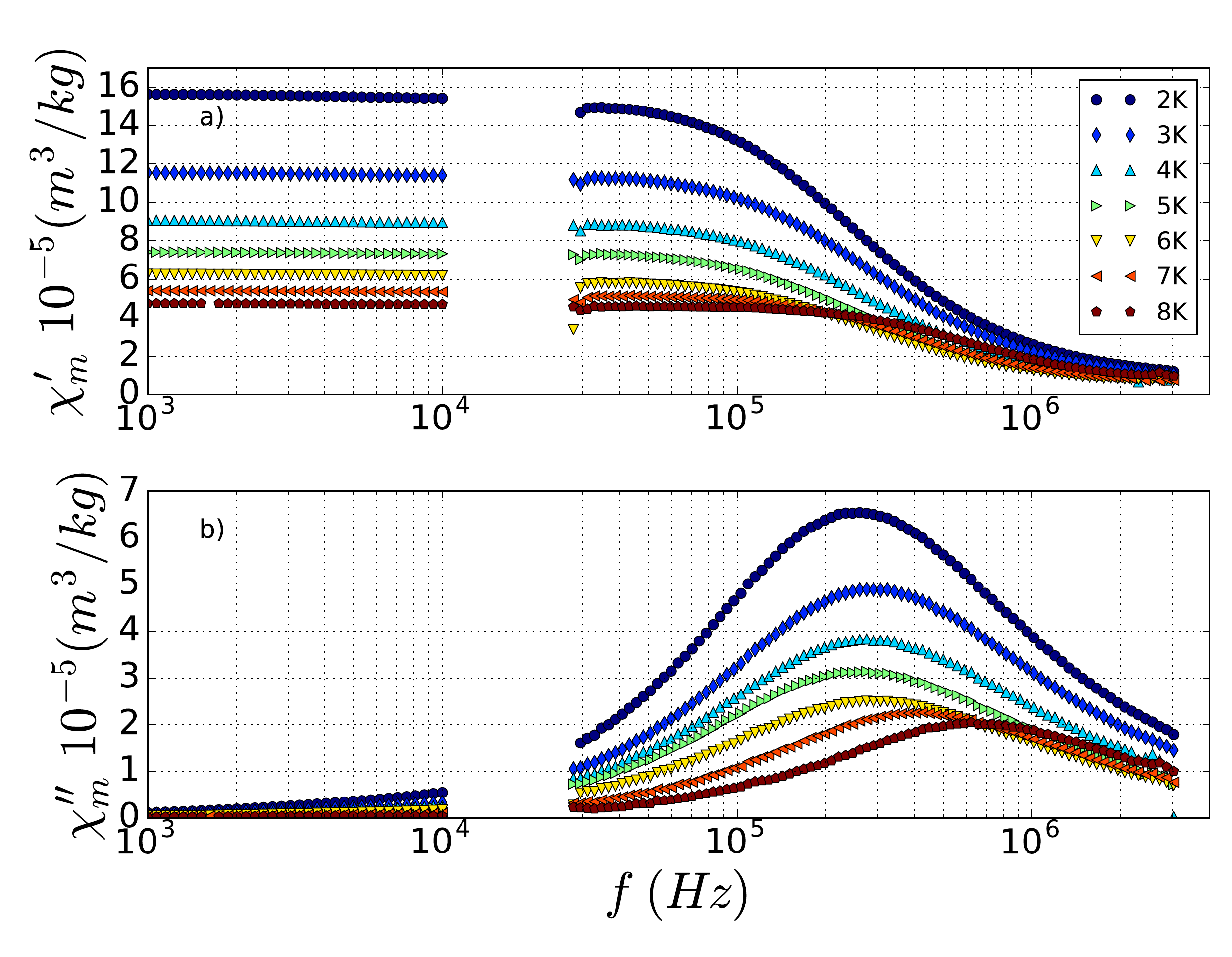}
\caption{\label{Sele} \textit{Example of mass ac-susceptibility data vs. frequency of $CdEr_2S_4$ at low temperatures. a) the in-phase component and b) the out-of-phase component. Frequencies $<10^4 Hz$ were measured by the QD-ACMS, frequencies $>10^4 Hz$ were measured by our system.}}
\end{figure}

Data taken on two further samples are presented to illustrate the performance of the system at low and high temperatures. The first sample is CdEr$_2$S$_4$, which is a dipolar spin ice material with a spinel structure. Spin ices are materials in which geometric frustration on the crystal lattice causes novel magnetic behavior, including emergent quasi-particles and a residual entropy in the ground state \cite{Ramirez1999}. High frequency susceptibility measurements on CdEr$_2$S$_4$ with the detailed in-phase and out-of-phase components  shown in figure~\ref{Sele}a and b were performed as part of a recent study into this material \cite{Gao2018}.

Magnetic nanoparticles suspended in water were also used to illustrate the ac susceptometer working across a large temperature range. Magnetic nanoparticles can be used in biomedical applications such as magnetic particle imaging and magnetic fluid hyperthermia \cite{Ferguson2013,Q.A.PankhurstS.K.JonesJ.Dobson2003}. We measured an iron oxide based nanoparticle system produced by nanoPET Pharma GmbH, FeraSpin XS, which have a hydrodynamic particle diameter in the range 10-20~nm \cite{Mbiotech}. The peak in the imaginary part is expected to be broad \cite{Wetterskog2017}, which allows us to illustrate the susceptometer measuring over a large temperature range. 

\begin{figure}[!ht]
\centering
\includegraphics[width=0.8\linewidth,keepaspectratio]{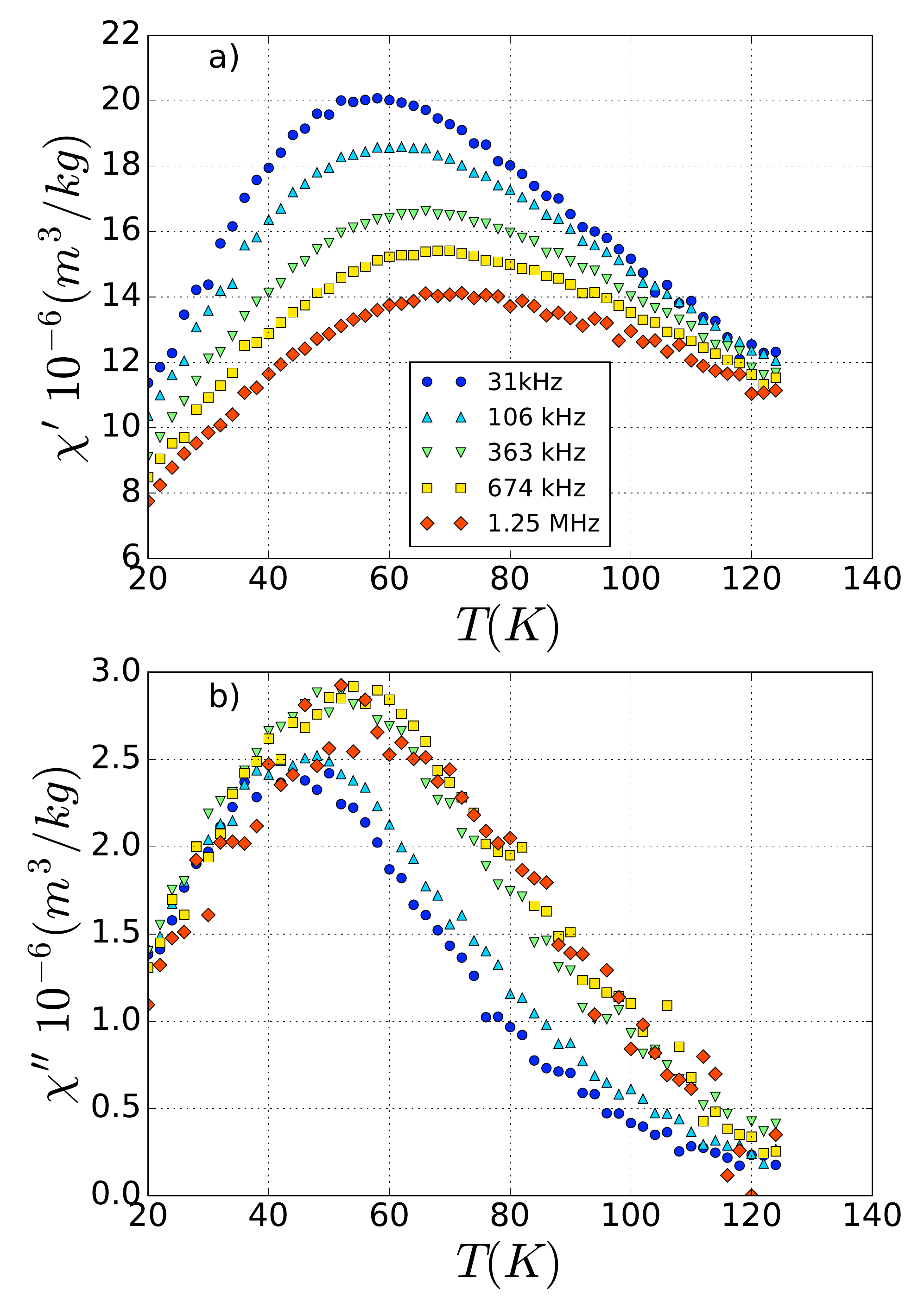}
\caption{\label{MNP} \textit{Mass ac-susceptibiltiy data vs temperature for five frequencies of the FeraSpin XS magnetic nanoparticles. a) In-phase component. b) Out-of-phase component.}}
\end{figure}

The suspension liquid was frozen by cooling to 250~K at atmospheric pressure with zero applied field, then waiting for an hour before purging the sample chamber and cooling to measurement temperatures. The magnetic nanoparticles were measured at frequencies ($f$=31 kHz, 106 kHz, 363 kHz, 674 kHz and 1.25 MHz) across the temperature range 20-124~K and the result can be seen in figure~\ref{MNP}. The maximum in-phase and out-of-phase components at a specific temperature, as shown in figure~\ref{MNP}, is because the N\'eel relaxation time becomes comparable to the measuring time. 

\begin{figure}[!ht]
\centering
\includegraphics[width=\linewidth,keepaspectratio]{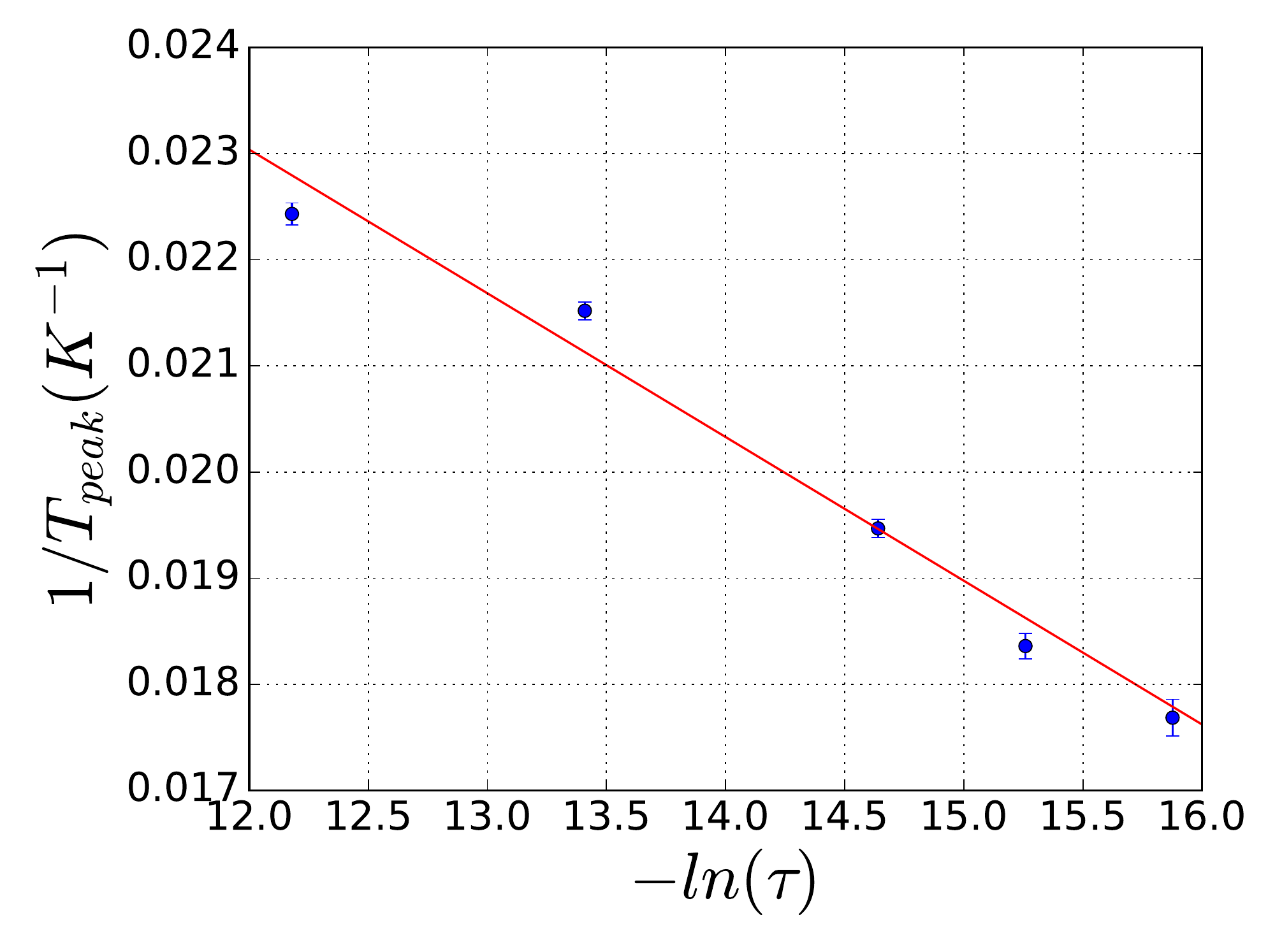}
\caption{\label{MNPA}\textit{Results of our analysis of the magnetic nanoparticle data in fig.~\ref{MNP} showing the inverse of the peak location in $\chi ''$ against the logarithm of the corresponding relaxation time for the relevant measurement frequency.}}
\end{figure}

An Arrhenius equation (eq.~\ref{NARR}) which assumes coherent magnetization reversal in the magnetic core and non-interacting cores subject to small magnetic fields was used to fit the determined peak temperatures at different relaxation times:

\begin{equation}\label{NARR}
\tau(T)=\tau_0\exp\left(\frac{KV}{k_BT}\right).
\end{equation}

For single core non-interacting nanoparticles this is a reasonable assumption~\cite{Bogren2015}. $\tau_0$ is often referred to as the attempt frequency and is usually in the range of 10$^{-9}$ s and 10$^{-10}$ s for iron oxide based nanoparticles, $K$ is the magnetic anisotropy constant, V is the magnetic core volume and $k_B$ is the Boltzmann constant. Figure~\ref{MNPA} shows the inverse of the peak locations in temperature of $\chi ''(T)$ against the corresponding intrinsic relaxation time associated with the measurement frequency according to $\tau = 1/2 \pi f$. Peak locations and associated errors were estimated from Lorentzian fits to the data.
By using the magnetic anisotropy constant suggested by \textit{Wetterskog et al}~\cite{Wetterskog2017} ($K=2.8\times10^{4} J/m^{3}$) for this nanoparticle system we find the mean core diameter to be ($8.86\pm0.09$) nm and $\tau_0=(2.50\pm2.35)\times10^{-13}$ s. Our Arrhenius analysis has found that $\tau_0$ is of the same order of magnitude as estimated from the Arrhenius plot by \textit{Wetterskog et al}.
The mean core diameter we find is slightly larger than reported by \textit{Wetterskog et al} using transmission electron microscopy analysis ($\sim$6 nm) but there is a good resemblance of the FeraSpin XS core diameter with recently reported data~\cite{Bender2018} using X-ray diffraction (XRD), neutron powder-diffraction (ND) and small angle x-ray scattering (SAXS), ($\sim$9 nm). The different values of the core size can be due to different size weightings using different analysis techniques~\cite{Bender2018}. The low value of $\tau_0$ can be due to magnetic interactions between the magnetic cores.

\section{Summary}
We have built an ac susceptometer capable of measuring the in-phase and out-of-phase ac susceptibilities at frequencies up to 3.5~MHz, across the temperature range 2-300~K, and in dc fields up to 9~T. This capability allows easy measurement of bulk samples in a range of temperatures, magnetic fields and frequencies previously very difficult to observe. Careful design of the coil winding has allowed us to minimize parasitic capacitance and inductance which, along with the calibration routines described, allow us to maintain stable precise measurements to high frequency with good sensitivity. We have reported on several example measurements performed on samples with different magnetic properties using the high frequency ac susceptometer illustrating its range and stability. Ac susceptibility measurements on MgB$_2$ and Dy$_2$Ti$_2$O$_7$ have shown that heating from the coil set is negligible and that the sample temperature is well known. Measurements performed on CdEr$_2$S$_4$ illustrate almost the full frequency range available and measurements on an iron oxide based magnetic nanoparticle system show functionality over a wide temperature range. 

\section{Acknowledgements}
SRG thanks ESPRC EP/L019760/1 for funding.The authors would like to thank T. Fennell and O. Zaharko for samples.

\bibliography{library2}{}
\bibliographystyle{unsrtnat}

\end{document}